\documentclass[preprint,showpacs,prc,aps,showkeys]{revtex4-1}
\usepackage{graphicx}% Include figure files
\usepackage{dcolumn}% Align table columns on decimal point
\usepackage{bm}% bold math
\usepackage{hyperref}% add hypertext capabilities

\begin{document}

\preprint{APS/123-QED}

\title{Systematics of Anti magnetic rotation in even-even Cd isotopes }

\author{Santosh Roy$^1$
}\altaffiliation[Also at ]{Saha Institute of Nuclear Physics. 1/AF Bidhannager, Kolkata 700 064, India}
\author{ S. Chattopadhyay$^2$}
\author{Pradip Datta$^{3}$}
\altaffiliation[Also at ]{Ananda Mohan College, 102/1 Raja Rammohan Roy Sarani, Kolkata 700 009, India}
\author{S. Pal$^{2}$}
\altaffiliation[Presently at]{ IRFU, CEA Saclay, 91191, Gif-sur-Yvette, France}
\author{S. Bhattacharya$^2$}
\author{R. K. Bhowmik$^4$, A. Goswami$^2$, H. C. Jain$^5$, R. Kumar$^4$, S. Muralithar$^4$, D. Negi$^4$, R. Palit$^5$, R. P. Singh$^4$}

\affiliation{$^1$S. N. Bose National Centre for Basic Sciences. Block JD, Sector III, Saltlake City, Kolkata 700098, India}

\affiliation{$^2$Saha Institute of Nuclear Physics, 1/AF Bidhannager Kolkata, 700 064, India}

\affiliation{$^3$iThemba Labs,P.O. Box 722 Somerset West 7129, South Africa.}

\affiliation{$^4$Inter University Accelerator Center, Aruna Asaf Ali Marg, New Delhi 110 067, India}

\affiliation{$^5$Tata Institute of Fundamental Research, Homi Bhabha Road, Mumbai 
400 005, India}

\date{\today}

\begin{abstract}
The lifetimes for the high spin levels of the yrast band of $^{110}$Cd has
been measured. The estimated B(E2) values decrease with increase in angular
momentum. This is the characteristic of Anti magnetic rotation as reported 
in $^{106,108}$Cd. However, alignment behavior of $^{110}$Cd is completely 
different from its even-even neighbors. A model based on classical particle 
plus rotor has been used to explore the underlying
systematics and develop a self consistent picture for the observed behavior of
these isotopes. 
\begin{description}

\item[PACS numbers] 21.10.Re, 21.10.Tg, 21.60.Ev, 23.20.-g, 27.60.+j

\end{description}
\end{abstract}

\maketitle

A deformed nucleus exhibits quantum rotation about an Principal axis, which is perpendicular to it's axis of symmetry. This collective behavior generates a band structure with the level energies (E) proportional to I(I+1) \cite{Bohr} , where I is the angular momentum. This behavior is very similar to the rotation of a diatomic molecule. But a deformed nucleus can also generate angular momentum by aligning nucleons along the rotational axis. These alignments (termed as band crossing) produce abrupt increase in angular momentum and breaks the smooth I(I+1) behavior. Thus, in a well deformed nucleus the angular momentum gets generated through an interplay of collective and single particle excitation. 

\indent It is evident from the above picture that if one considers a nearly spherical nucleus  with a substantial number (four or more) of valence nucleons then, the total angular momentum may almost fully be generated by the valence nucleons. 
A very special situation arises near certain shell closures where the neutrons (protons) just above a shell closure (particles) have their angular momentum along the rotational axis (\textbf{$\bm{j}_p$}) while for the protons (neutrons) just below the closure (holes) the angular momentum is along the symmetry axis (\textbf{$\bm{j}_h$}). 
Thus, the resultant angular momentum is tilted with respect to both symmetry and rotational axes \cite{fraun1, fraun2}. 

\indent In this situation, the band head spin is generated by the 
perpendicular coupling of $\bm{j}_p$ and $\bm{j}_h$. The higher spin 
states of the band are generated by gradual closing of these two 
vectors around their resultant which resembles the closing of a pair of 
shear and the excitation energy along the band increases due to the 
potential energy associated with the reorientation of the two vectors. 
Thus, this shears model leads to a smooth monotonic increase of energy 
with angular momentum and was proposed by Clark and Macchiavelli \cite{machia1, rmc1, rmc2}. 
In this semi-classical geometric model, the only important degree of 
freedom is the angle between $\bm{j}_p$ and $\bm{j}_h$ and is known 
as the shears angle, $\theta$. The highest spin state for such a 
shear structure is $\bm{j}_p + \bm{j}_h$ and is achieved for $\theta = 0$. 
This shears structure of particle and holes leads to a large magnetic 
dipole moment$(\mu)$ which precesses around $\bm{I}$ and leads to a 
large M1 transition rate (B(M1)) for the shears band as it is 
proportional to $\mu_{\perp}^{2}$. However, as the shear closes 
($\theta$ decreases), $\mu_\perp$ decreases and the B(M1) 
rates also decrease with increasing angular momentum \cite{fraun1, fraun2, pd, datta2}.

\indent A large number of such bands with sequences of M1 transitions have been found in A$\sim$ 200, A$\sim$ 140 and A$\sim$ 100 mass regions. These bands are known as M1 or Shears band where the B(M1) rates exhibit the characteristic fall. In few cases band crossing in Shears band have also been found and the observed features have been well described by the semi-classical model~\cite{santosh, aks, cooper}.

\begin{figure}
\includegraphics*[scale=0.7]{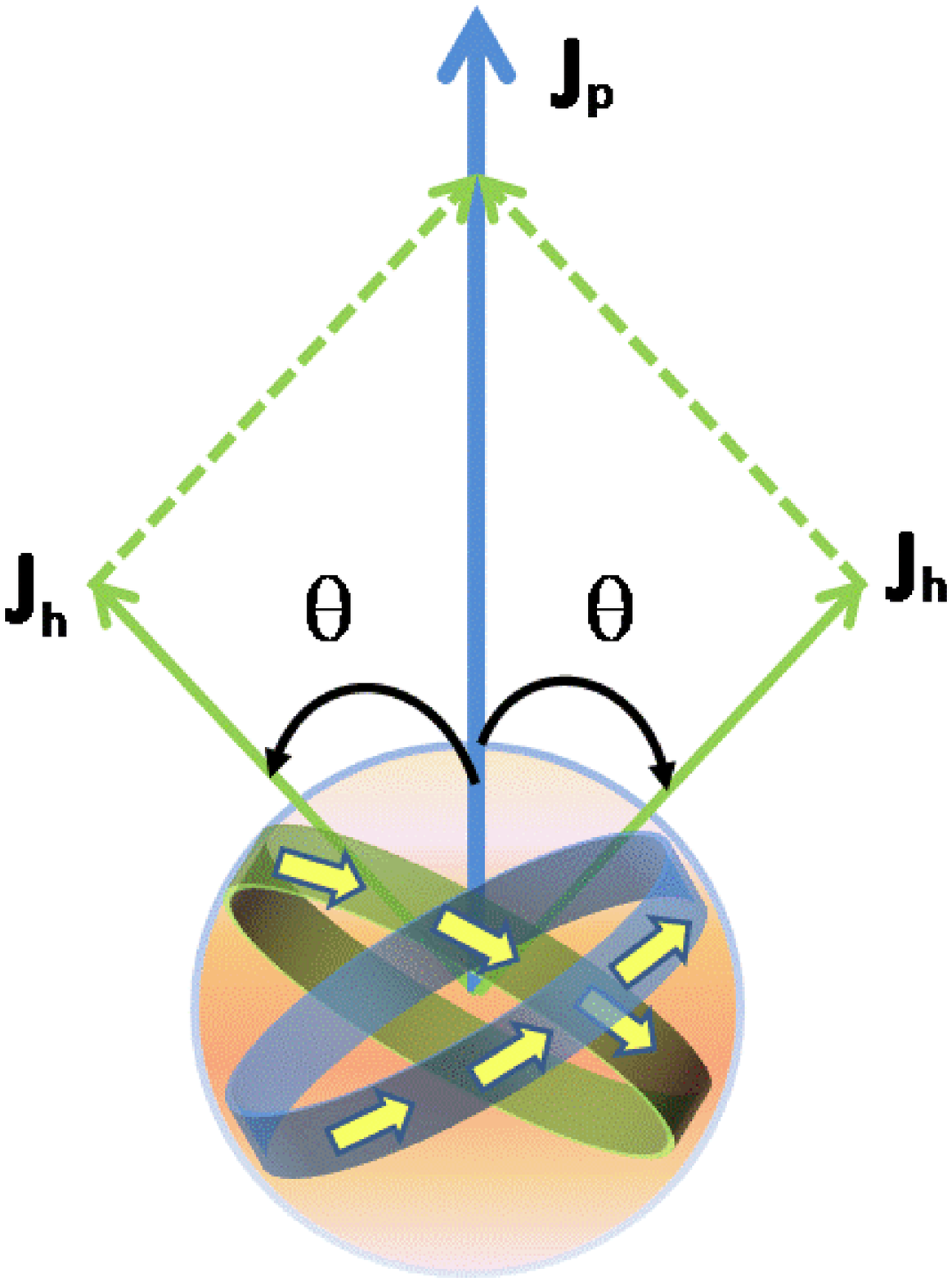}
\caption{\label{fig:fig1}~ Vectorial representation of Anti magnetic rotation, where, $\bm{I}$, $\bm{j_p}$ and $\bm{j_h}$ are the total, particle and hole angular momentum vectors, respectively. $\theta$ is the shears angle.}
\end{figure}

\indent An interesting consequence of the shear structure has been pointed out by Frauendorf \cite{fraun1, fraun2}. It is possible to have a symmetric double shear structure with respect to the rotational axis. The vector diagram for angular momentum is shown in Fig.~\ref{fig:fig1}, where the particle and hole angular momentum vectors,  
$\bm{j_p}$ and $\bm{j_h}$, are along the rotational and symmetry axes respectively, $\theta$ is angle between the particle-hole vectors and 2$\theta$ is the angle between the two hole vectors. The higher angular momentum states in this scenario, will be generated by closing the shear angle and is represented by
\begin{equation}
I=\ j_p + 2 j_h \cos \theta
\label{eqn:eqn1}
\end{equation}

\indent In this specific case, the $\pi$-rotational symmetry is restored and thus, the band will consist of electric quadrupole (E2) transitions. In the semi-classical model, the two perpendicular components of the dipole moments cancel each other due to the symmetry of the structure which leads to the absence of M1 transitions. This coupling scheme has been termed as Anti Magnetic Rotation (AMR) due to it's similarity with anti ferro-magnetism, where the dipole moment of one of the sub-lattice is aligned in the opposite direction to that of the other half which leads to the absence of magnetization. In this model, the electric quadrupole transition rate (B(E2)) is proportional to ${\sin^4 \theta}$. Therefore, as the two shears close symmetrically, the B(E2) rates will show a characteristic drop with increasing angular momentum. This behavior of B(E2) values is the crucial experimental signature which distinguishes a AMR band from a band arising due to collective rotation.

\indent This phenomenon has only been observed in the yrast bands of two nuclei namely, $^{106, 108}$Cd, which lie near the N=50 shell closure \cite{simons, pd1, simons2}.
 The properties of the AMR in these two bands are very similar. In both these cases, the band head spin is I= $16\hbar$ and the associated configuration is $(\pi {g_{(9/2)}}^{-2})_{J=0} \otimes \nu {{(g_{7/2})}^2}_{J=6}\ {{(h_{11/2})}^2}_{J=10}$. The higher spin states of these bands are formed through gradual alignment of the proton holes. Thus, the spin at the band termination is I= $24\hbar$ and the configuration is $(\pi {g_{(9/2)}}^{-2})_{J=8} \otimes \nu {{(g_{7/2})}^2}_{J=6}\ {{(h_{11/2})}^2}_{J=10}$. In $^{106}$Cd, the AMR band extends till $26\hbar$ and the experimental data suggest that the deformed core contributes $\sim\ 2\hbar$ angular momentum \cite{simons}. 

\indent The measured B(E2) rates for the states of AMR band in the two cases also match within error bars and shows the characteristic fall. These values were well reproduced by the geometric model of shear mechanism \cite{simons} where

\begin{equation}
B(E2)=\ \frac{15}{32 \pi} (eQ_{eff})^2\ {sin^4  \theta}
\label{eqn:eqn2}
\end{equation}
and in both cases $(eQ_{eff})=\ 1.1\ eb$ \cite{simons, pd1} was used.

\begin{figure}
\includegraphics*[scale=0.7]{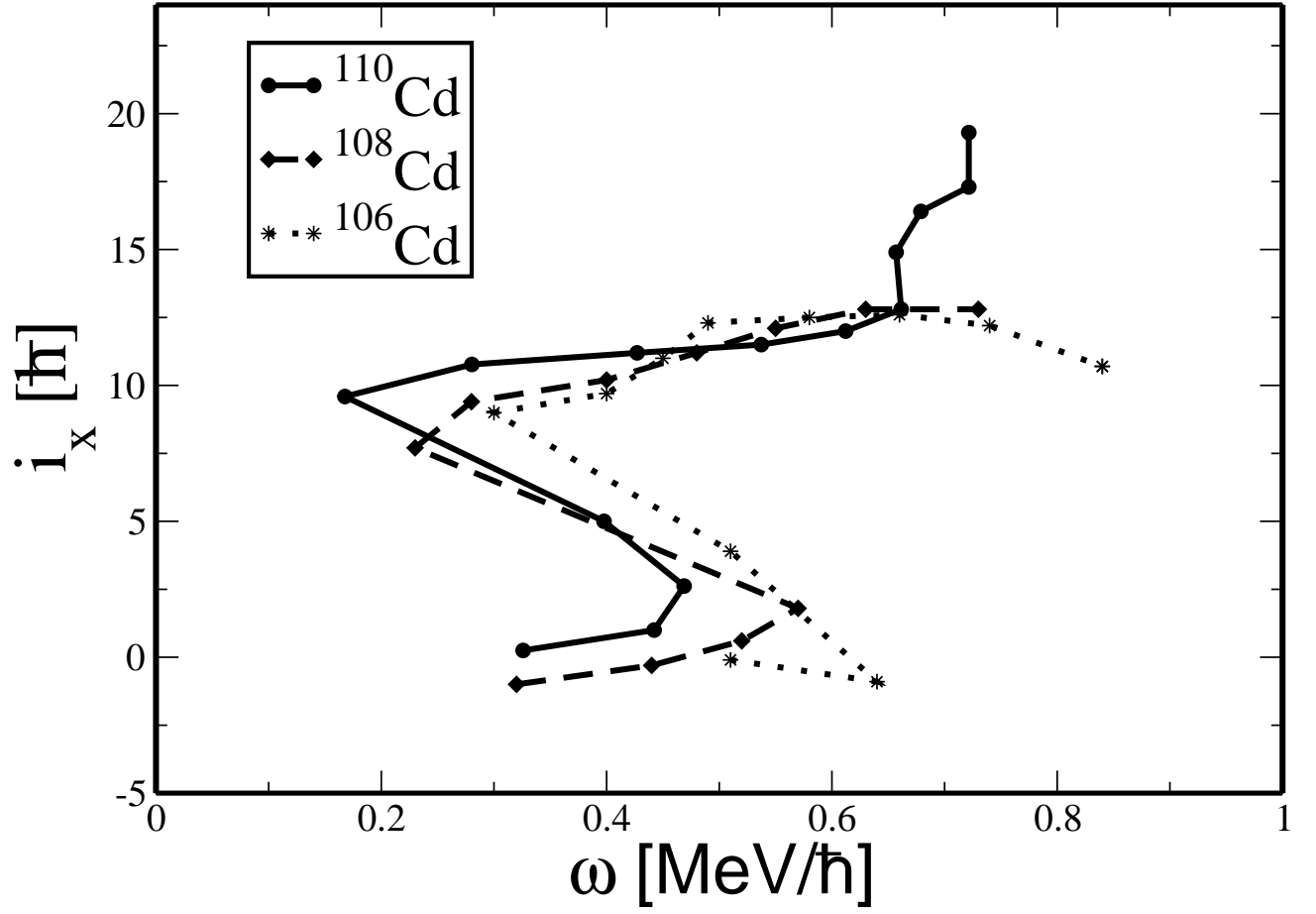}
\caption{\label{fig:fig2}~ Experimental aligned angular momentum for yrast band in $^{110}$Cd (solid line), $^{108}$Cd (dashed) and $^{106}$Cd (dotted), using the Harris parameters $J_0$= 5 $\hbar^2$/MeV and $J_1$= 15$\hbar^4$/Mev$^3$. }
\end{figure}

\indent This close similarity of AMR bands in $^{106,108}$Cd indicates a natural expectation of a similar structure in the high spin states of the yrast band of $^{110}$Cd. 
However, the high spin behavior of $^{110}$Cd is completely different from its 
even-even neighbors as was established by S. Juutinen {\sl et al.} \cite{juut}. This has been demonstrated in Fig.~\ref{fig:fig2}, where the aligned angular momenta ($i_x$) of $^{106,108,110}$Cd have been plotted against the rotational frequency ($\hbar \omega$). In $^{106,108}$Cd \cite{regan, thors}, a sharp backbend is observed around $\hbar \omega \sim 0.4$ MeV due to alignment of $h_{11/2}$ neutron which leads to an large alignment 
gain of $\sim 10 \hbar$. Immediately after the back bend, there is a again a small alignment gain due to alignment of two further $g_{7/2}$ neutrons. Beyond $\hbar\omega \sim\ 0.5$ MeV, a very small and slow alignment gain is observed which is due to the alignment of the two proton holes. This is the domain of frequency where AMR has been observed. In $^{110}$Cd, the $h_{11/2}$ alignment occurs at very similar frequency, but there is no further alignment gain till 0.6 MeV. Thus, the $g_{7/2}$ neutron alignment observed in this frequency range in $^{106,\, 108}$Cd, is absent in $^{110}$Cd. Beyond $\hbar\omega=\ 0.65$ MeV, there is a large alignment gain of $8\hbar$ which continues till the highest observed frequency of 0.72 MeV. 
This phenomenon is absent in $^{106,108}$Cd and no satisfactory explanation of such a large alignment gain at $\hbar \omega= 0.65$ MeV in $^{110}$Cd could be given from Cranking Shell Model calculations which assumes collective rotation \cite{juut}.

\indent The present work, reports the measurement of lifetimes of the high spin levels for the yrast band of $^{110}$Cd and aims to develop a consistent description of the observed systematics of even-even Cd-isotopes based on the semi-classical geometric model of AMR. The magnitude of the strength of the hole-hole interaction for a shear structure has also been estimated for the first time.

\indent The high spin states of $^{110}$Cd were populated through the reaction $^{96}Zr(^{18}O,4n)^{110}Cd$ at a beam energy of 70 MeV delivered by the 15UD Pelletron Accelerator \cite{gkm} at Inter University Accelerator Centre (IUAC), New Delhi. The $\gamma$-rays were detected in the Indian National Gamma Array (INGA) \cite{murali}, that consisted of 18 Compton suppressed Clover 
detectors, with two at $32^\circ$, four at $57^\circ$, four at $90^\circ$, four at $123^\circ$ and four at $148^\circ$ with respect to the beam axis. The target was made of 1 $mg/cm^2$ enriched $^{96}$Zr on 9 $mg/cm^2$ $^{206}$Pb backing. A total of 1.3 $\times$ $10^9$ $\gamma-\gamma-\gamma$ events were collected. 
The data was sorted to form a number of angle-dependent asymmetric $\gamma$-gated matrices with the gates on the ($2^+ \rightarrow\ 0^+$), ($4^+ \rightarrow\ 2^+$) and ($6^+ \rightarrow\ 4^+$) transitions of $^{110}$Cd using INGASORT \cite{bhow1} program. These matrices were constructed with $90^\circ$ detectors in one of axis and forward or backward detectors in the other axis for LINESHAPE analysis. The lineshapes were extracted using 335 keV ($10^+ \rightarrow\ 8^+$), 561 keV ($12^+ \rightarrow\ 8^+$) and 854 keV ($14^+ \rightarrow\ 12^+$) $\gamma$-gates on the $\gamma$-gated asymmetric matrices. 

\indent In the present work, the lineshape were observed above the I$^\pi$= 16$^+$ level and the lifetimes of these high spin levels were estimated using the LINESHAPE analysis code of Wells and Johnson \cite{john1}. This code was used to generate the velocity profile of the recoiling nucleus into the backing using Monte Carlo technique with a time step of 0.001 ps for 5000 histories. These profile were generated at 123$^\circ$, 148$^\circ$ and 57$^\circ$ where the clover geometry of the detectors had been incorporated. The detectors at $32^\circ$ ring were not considered since their energy resolutions were worse than others. The electronic stopping powers of Northcliff and Shilling corrected for shell effects were used for calculating the energy loss \cite{mil1}. 
%The detail procedure for lineshape fitting is described in ref[]

\indent The energies of $\gamma$ transitions and the side-feeding intensities
were used as input parameters for the lineshape analysis.
The side-feeding intensities were estimated from the intensity
profile obtained from the gated spectra at 90$^\circ$. The side feeding
into each level and feeding to the top most level of each band
was initially modeled as a cascade of five transitions with
a moment of inertia which was comparable to that of the
band of interest. The quadrupole moments of the side-feeding
sequences were allowed to vary which when combined with
the moment of inertia gave an effective side-feeding lifetime parameters for each level. For every observed lineshape,
in-band and side-feeding lifetimes, background parameters,
and the intensities of the contaminant peaks were allowed to
vary. For each set of the parameters the simulated lineshapes
were fitted to experimental spectra using $\chi^2$-minimization
routines of MINUIT \cite{james1} .

\indent The lifetime measurements were performed starting with
the top-most transition which was assumed to have 100$\%$ side
feed. The other parameters were allowed to vary until the
minimum $\chi^2$ was reached. The background and the stopped
contaminant peak parameters for best fit were then fixed. Thus,
the lineshape analysis for the top-most transition led to the
estimation of the effective lifetime for the top-most level. But, in $^{110}$Cd, the energies of the $\gamma$-transitions de-exciting the topmost level ($28^+$) and the next level ($26^+$) are the same namely, 1443 keV \cite{juut}. Thus, in the present case, only a effective lifetime for the top feed to $24^+$ level could be estimated which was 0.35(02) ps. The side feeding intensity at $24^+$ level was fixed to reproduce the observed intensity pattern at 90$^\circ$ with respect to the beam direction.
Since the lineshapes for 1323 keV (22$^+ \rightarrow$ 20$^+$) and 1314 (20$^+ \rightarrow$ 18$^+$) $\gamma$-transitions
were found to overlap in the lower gates, the 1323 keV lineshape was separately extracted in 1314 keV gate. This lineshape was fitted with a top feed lifetime equal to the effective lifetime of 24$^+$ level. This extracted lifetime for 22$^+$ level was kept fixed during the global fit where the other four levels were included. During the global fit, the in-band and side feeding lifetimes were allowed to vary and this procedure was repeated for 57$^\circ$, 123$^\circ$ and 148$^\circ$.
 The uncertainties in the lifetimes were derived from
the behavior of the $\chi^2$ fit in the vicinity of the minimum. The
level lifetimes were also measured in multiple $\gamma$ gates to avoid
the effect of any contamination in the lineshape. Thus, the final
values for the level lifetimes were obtained by taking averages
from the fits at the three angles and all gates. However, it should
be noted that the quoted errors do not include systematic
error in the stopping power values which may be as large
as 20$\%$. The measured level lifetimes and the evaluated B(E2) rates are listed in Table~\ref{tab:table1}.

\indent It is evident from the Table~\ref{tab:table1}  that the observed B(E2) rates in $^{110}$Cd exhibits a steady fall as a function of angular momentum beyond I$\sim$ 18$\hbar$. This behavior is the characteristic signature of AMR which has also been observed in $^{106,108}$Cd as seen from the values in Table~\ref{tab:table1} \cite{simons, pd1}. The highest observed angular momentum state in $^{106,108,110}$Cd is $26\hbar$, $24\hbar$ and $28\hbar$, respectively. Since the full alignment of the two $g_{9/2}$ proton holes can generate a maximum angular momentum of $8\hbar$, it may be assumed that the band head ($\theta=\ 90^\circ$) for AMR in these three cases are $18\hbar$, $16\hbar$ and $20\hbar$, respectively. 

\indent However, it is worth noting that the B(E2) values for $^{110}$Cd at 20$\hbar$ is lower by $\sim \ 25\%$ than the other two cases (at $16\hbar$ and $18\hbar$) which would imply that the effective quadrupole moment in Eq.~\ref{eqn:eqn2} is substantially lower ($\sim \ 50\%$) in $^{110}$Cd. This is in contradiction to TRS calculations which predicts nearly same deformation ($\beta_2 \sim 0.15$) for all the three isotopes. Thus, in $^{110}$Cd \cite{juut}, though the high spin states of the yrast band seems to originate from AMR, the alignment behavior and the observed B(E2) transition rates are completely different from it's immediate even-even neighbors.

\indent In order to explore any underlying systematics, we have used the classical analog of two-particle-plus-rotor model which was first proposed by Clark and Macchiavelli \cite{machia1, rmc1, rmc2}. For the Cd-isotopes, the two symmetric shears are formed between ${j_h}^{(1)}$ = ${j_h}^{(2)}$ = $j_\pi$ and $j_p= j_\nu= \ aj_\pi$, and the band head for this double shear structure is at $\theta_1$= $\theta_2$= $\theta$= $90^\circ$ (see Fig.~\ref{fig:fig1}). The higher angular momentum (I) states are formed by gradual closure of the two proton blades around ${j_\nu}$ and the maximum angular momentum generated in this scheme will be (${j_\nu}$ + $2{j_\pi}$ -1) for $\theta= 0$.

\indent A repulsive interaction of the form $V_{\pi \nu}\ P_2(cos \theta)$ between
 the particle-hole blades of a shear has been successful in describing the 
observed properties of the shears band in different mass regions \cite{rmc2}, where 
$V_{\pi \nu}$ is the interaction strength per pair. It has been 
argued by Macchiavelli {\sl et al.} \cite{machia1}  that such an interaction between the 
blades may be mediated through the core by a particle-vibrational coupling involving a 
quadrupole phonon \cite{machia2}. The magnitude of this interaction has been found to be around 
300 keV and 550 keV for A $\sim$ 200 and A $\sim$ 110 mass regions, respectively \cite{rmc1, santosh}. 
Thus, it has been experimentally established that this interaction scales as 1/A. 
The hole-hole (particle-particle) attractive interaction ($V_{\pi \pi}$) has been 
assumed to be of the same form with the additional boundary condition that it 
vanishes for $\theta=\ 0$ \cite{sugawara}. This condition implies that in the present case, the 
particle-particle attractive interaction will be absent. It is to be noted that 
the magnitude of this interaction has not been estimated since no Shears band 
with hole-hole (particle-particle) blades has been observed. Thus, in this 
classical particle-rotor model the energy E(I) is given by, 
\begin{equation}
E(I)=\ \frac{(\bm{I-j_p - j_h})^2}{2\Im}+\ V_{\pi \nu}(\frac{3 {\cos^2 \theta}-1}{2}) +V_{\pi \nu}(\frac{3 {\cos^2 (-\theta)}-1}{2}) - \ \frac{V_{\pi \pi}}{n} (\frac{3 {\cos^2 (2\theta)}-3}{2})
\label{eqn:eqn3}
\end{equation}
\noindent where the first term is the rotational contribution and the rest of the terms are the shears contribution. The functional form of $V_{\pi \pi}$ is chosen to respect the boundary condition and `n' is the actual number of neutron-proton (particle-hole) pairs for a given single particle configuration. The shears angle ($\theta$)
has been evaluated from Eq.~\ref{eqn:eqn1}. The corresponding angular momentum 
(I) can be evaluated by imposing the energy minimization condition as  function of 
$\theta$ and is given by

\begin{equation}
I=\ aj+\ 2jcos\theta+\frac{3 \Im V_{\pi\nu} cos \theta}{j}-\frac{6 \Im V_{\pi \pi} cos 2\theta\  cos\theta}{nj}
\label{eqn:eqn4}
\end{equation}

\indent For a band, originating solely from AMR, the moment of inertia of the core ($\Im$) is set to zero \cite{machia1, rmc2} and the expression for angular momentum reduces to Eq.~\ref{eqn:eqn1}. For such a band, the rotational frequency ($\omega_{sh}$) can be computed through ($\frac{dE_{sh}}{d\theta}/ \frac{dI_{sh}}{d\theta}$) and is given by,

\begin{equation}
\omega_{sh}=\ (3 V_{\pi\nu}/j)cos\theta -(6V_{\pi\pi}/nj)cos2\theta \ cos\theta
\label{eqn:eqn5}
\end{equation}

These expressions have been used to compute the total angular momentum as a function of 
frequency (I($\omega$) plot) AMR and compared with the experimental data.

\indent For $^{108}$Cd, the states between 16$\hbar$ and 24$\hbar$ was found to originate due to AMR, and the corresponding single particle configuration is 
$\pi {g_{9/2}}^{-2} \otimes  \nu [{h_{11/2}}^2, {g_{7/2}}^2]$. 
This configuration has eight neutron-proton pairs (n=8), and $j_\nu$= 16, 
$j_\pi$= 4.5 i.e, a= 3.55. Fig.~\ref{fig:fig4}(a), shows the comparison of 
theoretical and experimental routhians where the calculated frequency at the 
band head is shifted by 0.5 MeV in order to match the experimental observation. 
This is due to the fact that the proton alignment in $g_{9/2}$ starts around $\hbar \omega= 0.5$ MeV as seen from the alignment plot (see Fig.~\ref{fig:fig2}), which marks the start of Anti magnetic rotation.
The calculation has been performed for $V_{\pi \nu}$= 600 keV which is consistent 
with the systematics of the mass-range. Since no estimate of $V_{\pi \pi}$ is 
available, calculations have been performed for different values $V_{\pi \pi}$. It has been found that the calculated values give 
a good description of the experimental I($\omega$) behaviour when $V_{\pi \pi}$ is varied between 0 to 200 keV. This in-sensitiveness
 may be attributed to the fact that there is only one hole-hole shear compared to eight particle-hole shears in $^{108}$Cd. However, for $V_{\pi \pi}$= 300 keV (dash line in Fig.~\ref{fig:fig4}(a)), the 
theoretical I($\omega$) deviates from experimental data at high spins. 

\indent The B(E2) values have been calculated using Eq.~\ref{eqn:eqn2}. for $eQ_{eff}=1.1 \ eb$ \cite{simons, pd1}, where $\theta$ and total angular momentum (I) are related by 
Eq.~\ref{eqn:eqn1}. The calculated (solid line) and experimental values have been plotted in 
Fig.~\ref{fig:fig4}(b), which shows a good agreement. Thus, the observed features of the AMR band in $^{108}$Cd has been well described by the present model. 

\indent The situation is very similar for $^{106}$Cd, except $j_\nu$= 18 has been 
used, which is consistent with the assumption that $\sim\ 2\hbar$ comes from the core contribution\cite{simons}. The calculated and observed I($\omega$) plots and B(E2) values are shown in figure~\ref{fig:fig5} (a) and (b), respectively. The agreement is again satisfactory for $V_{\pi \nu}$= 600 keV and $V_{\pi \pi}$ can be varied between 0 and 200 keV. 

\indent The situation changes in $^{110}$Cd. The expected configuration for the high spin states is $\pi {g_{9/2}}^{-2}\  \otimes\  \nu{h_{11/2}}^2$, since the alignment plot does not support the alignment of $g_{7/2}$ neutrons. Thus, for this configuration there are four neutron-proton combinations (n=4). For the calculation, the band head has been assumed to be 12$\hbar$, since the $h_{11/2}$ neutron alignment takes place around 10$\hbar$ which is essential for the formation of the double 
shear structure. Such an assumption will imply that the maximum angular momentum 
generated by shears mechanism is $20 \hbar$ and the remaining 8$\hbar$ is generated by core rotation. Therefore, for $^{110}$ Cd, the angular momentum (I) is to be calculated using the
 Eq.~\ref{eqn:eqn4} and the frequency $\omega$ will be given by,
\begin{equation}
\omega= \omega_{rot}-\omega_{sh}
\label{eqn:eqn6}
\end{equation}

\noindent where, $\omega_{sh}$ is given by Eq.~\ref{eqn:eqn5} and $\omega_{rot}$ = $\frac{1}{2\Im}(2I+1)$ is the core rotational frequency, where $\Im$ has been assumed to be
 19.2 MeV$^{-1} \hbar ^2$ which corresponds to half the rigid rotor value. It is interesting to note that this
value gives the same slope as the experimental routhian in the frequency interval of 0.4 to 0.6 (shown as the dot dashed line in Fig.~\ref{fig:fig6}(a)). The difference in the values along the angular momentum axis is due to the contributation of the two aligned $h_{11/2}$ neutrons.
  The relative negative sign takes into 
account the fact that in the present case, the angular momentum is generated both 
through collective and shear mechanism, Thus, a given angular momentum state will 
be formed at a lower energy (i.e. frequency) as compared to that due to pure 
collective rotation. For this calculation $V_{\pi \nu}$= 600 keV has been 
fixed since this value has been established from the systematics of $^{106, 108}$Cd.
 Thus, $V_{\pi \pi}$ is the only free parameter in the calculation of theoretical 
I($\omega$) plot which has been calculated for 
\begin{equation}
\omega=\ 0.026(2I +1)-\frac{3 V_{\pi \nu }}{j}cos \theta + \frac{1.5 V_{\pi \pi}}{j} cos2 \theta \ cos\theta
\label{eqn:eqn7}
\end{equation}

\indent The calculated values have been plotted as dotted, solid and dashed lines in Fig.~\ref{fig:fig6}(a) for $V_{\pi \pi}$= 0, 150 and 300 keV, respectively. It is evident from the figure that the experimental I($\omega$) beyond 0.65 MeV is well reproduced for $V_{\pi \pi}$= 150 keV and has a definative effect on the curvature of the routhian. The effect of $V_{\pi \pi}$ becomes appreciable in $^{110}$Cd since, there is one hole-hole shear combination and four particle-hole combinations, while in $^{106,\ 108}$Cd there are eight. Thus, the present calculation suggests that if the hole-hole interaction is assumed to have a functional form of $P_2(cos\theta)$, then the strength of this attractive interaction is around 150 keV for A= 110 region.

\indent It is interesting to note that in the present model, $\theta= \ 90^\circ$ at
 I = 12$\hbar$ which is the band head and $\theta=\ 65^\circ$ at 20$\hbar$. 
This is in contrast to a simplified picture where the band head for AMR could have 
been assumed to be 20$\hbar$. The B(E2) values have been calculated in both the 
scenario for $eQ_{eff}=\ 1.1 \ eb$ and plotted in Fig~\ref{fig:fig6}(b). The dashed 
line assumes a band head of $20 \hbar$ ($\theta= \ 90^\circ$) for AMR
 and is evident that this scenario 
fails to reproduce the observed B(E2) values. The rotation+shear picture, on the 
other hand, shows a good agreement with the observed values. Thus, the present 
work indicates that the high spin states of $^{110}$Cd originate due to an 
interplay between collective core rotation and Anti magnetic rotation. This 
scenario, gives a good description of both the experimental I($\omega$) behaviour and the 
observed E2 transition rates.

\indent In the present work, the features of the AMR in 
$^{106, 108, 110}$Cd has been described within the common framework of a 
classical particle-rotor model, where the particle-hole and hole-hole residual
 interactions have been assumed to be of $P_2$-type force. In all the three cases
 the strength of the repulsive particle-hole interaction has been found 
to be $\sim$ 600 keV which is consistent with the systematics of this mass 
range. The effect of attractive hole-hole interaction is weaker in case 
of $^{106,\, 108}$Cd. However, it's strength can be 
estimated in $^{110}$Cd since the curvature of the calculated routhian has a 
definitive dependence on $V_{\pi \pi}$ and found to be $\sim$150 keV. It is 
to be noted that this is the first instance where the strength of the hole-hole 
interaction has been estimated in case of a shears structure. The theoretical calculations also give a good description of the observed B(E2) rates in the three
Cd-isotopes. 

\indent Thus, the present study indicate that the high spin 
states of the yrast band of $^{106,\, 108}$Cd originates due to Anti magnetic rotation while those for $^{110}$Cd originates due to both collective and AMR. 
This seems to be the reason for the distinctive behavior for the high spin states of the yrast band of $^{110}$Cd.

\indent The authors would like to thank all the technical staff of the Pelletron facility at IUAC, New Delhi, for smooth operation of the machine. We would also like to thank Professor John Wells for providing the lineshape analysis package. This work was partly funded by the Department of Science and Technology, Government of India
(No. IR/S2/PF-03/2003-I).

\newpage

\begin{figure}
\includegraphics*[scale=0.7]{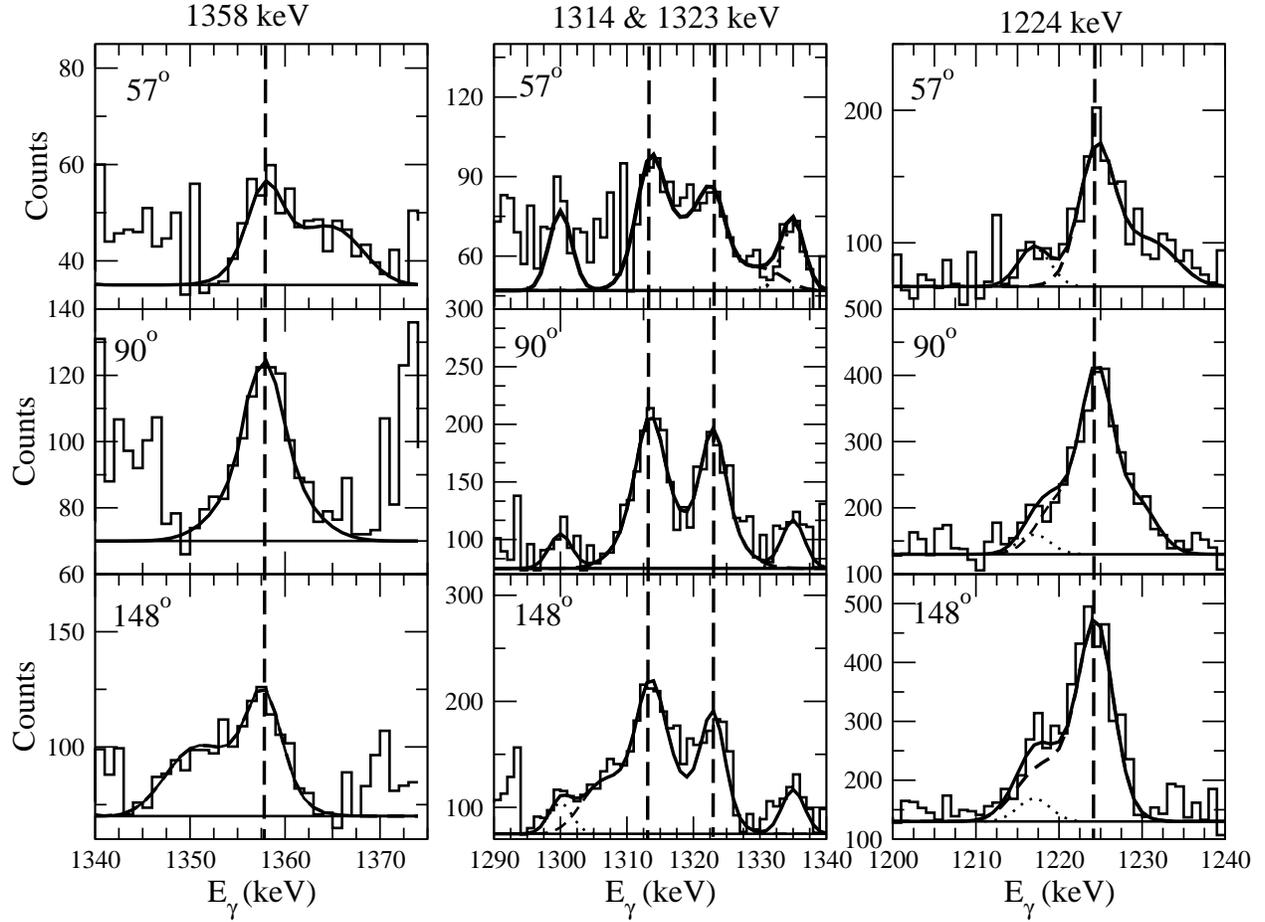}
\caption{\label{fig:fig3}~ Experimental data and associated line shape fits for the 1358, 1314, 1323 and 1224 keV transition in band 7 of $^{110}$Cd \cite{juut}. The angles for which the linespahes are fitted are shown in the left-hand side of the plot. The contaminant peaks are shown by the dotted lines. The unshifted $\gamma$-rays energy for each transition is marked by the vertical dashed line.}
\end{figure}

\begin{figure}
\includegraphics*[scale=0.7]{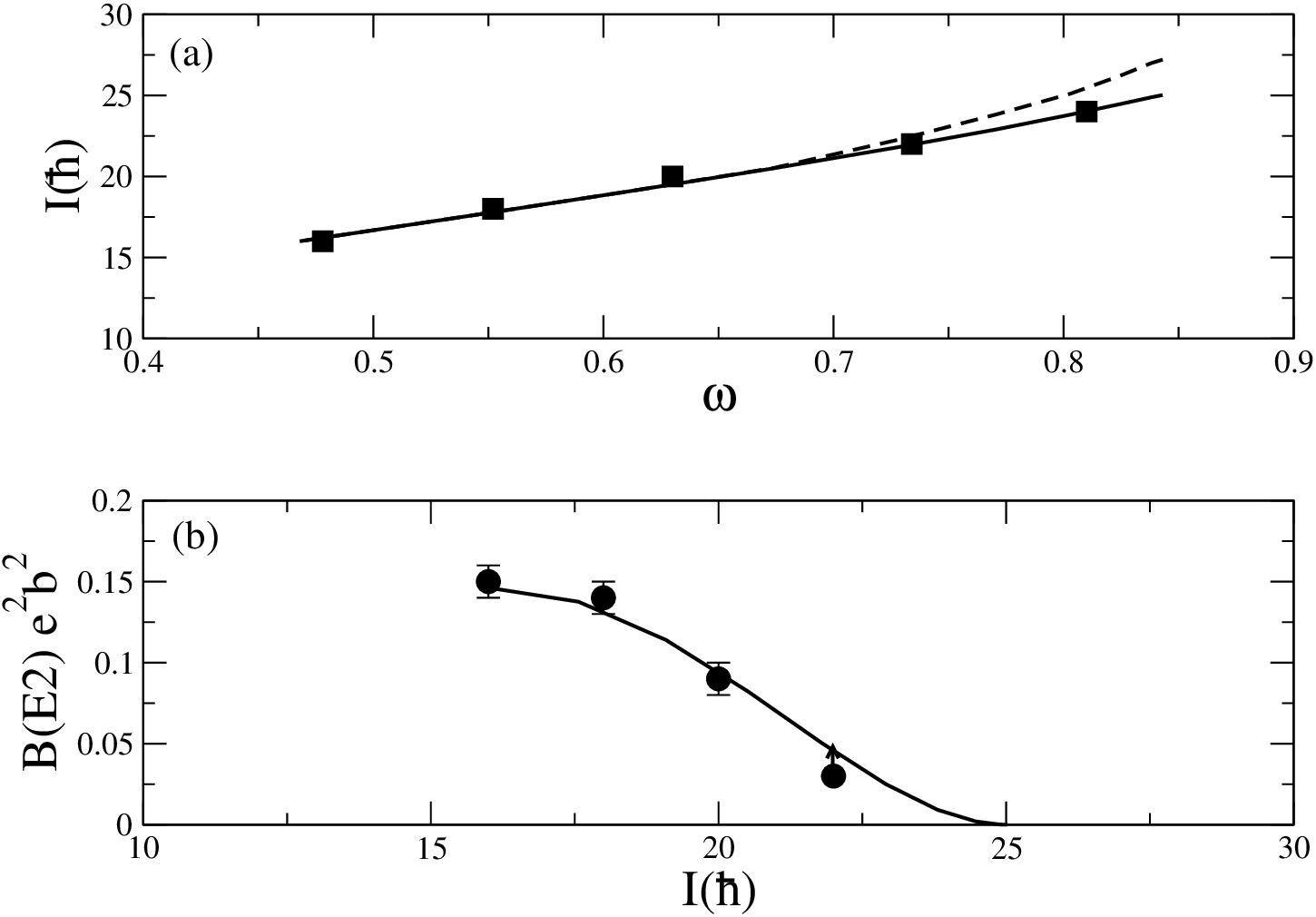}
\caption{\label{fig:fig4}~ The observed I($\omega$) (a) and B(E2) rates (b) in $^{108}$Cd. The lines represents the calculated values using the classical particle plus rotor model. The solid and dashed lines in (a) are the calculated values for $V_{\pi \pi}$= 150 and 300 keV, respectively.}
\end{figure}

\begin{figure}
\includegraphics*[scale=0.7]{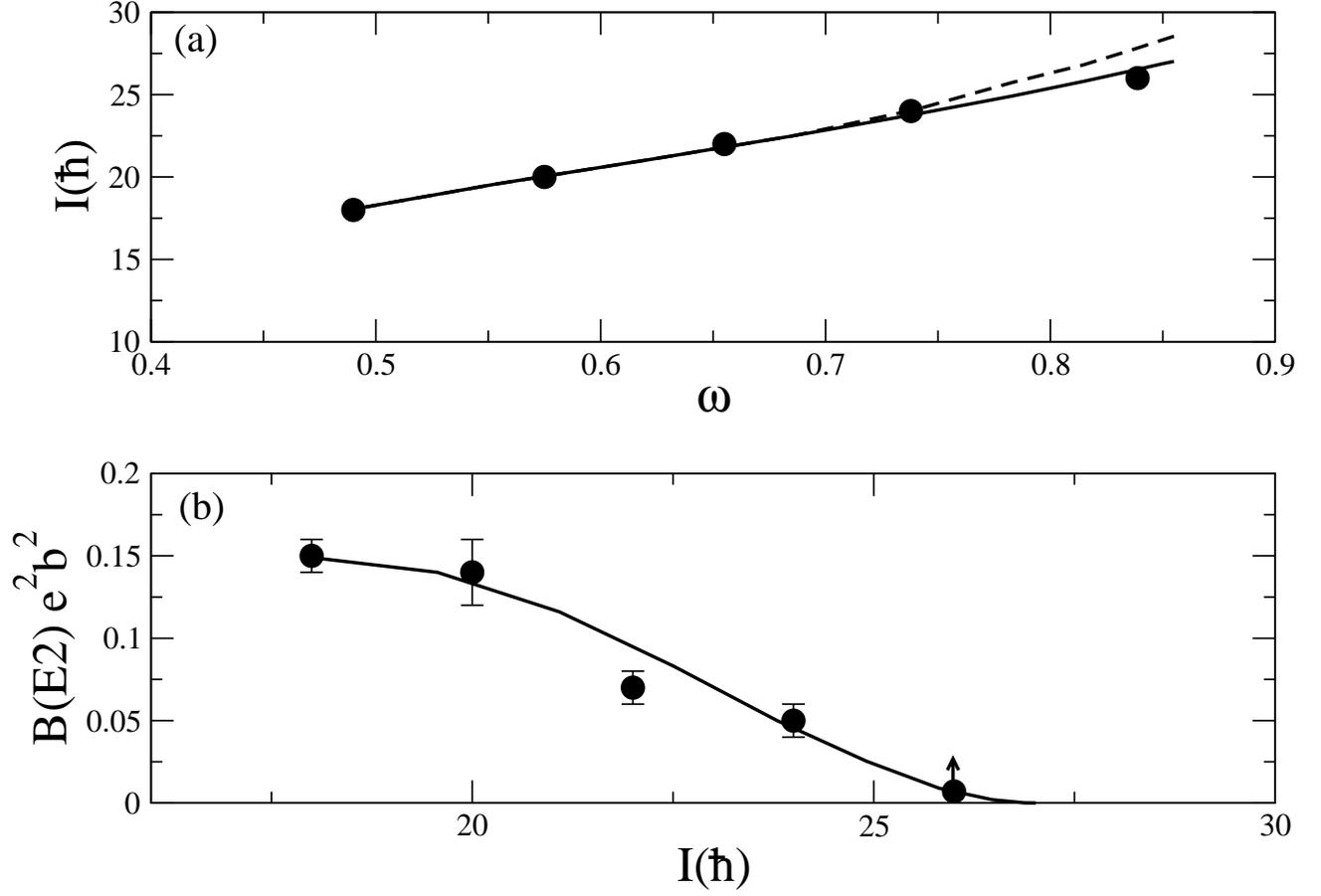}
\caption{\label{fig:fig5}~ The observed I($\omega$) plot (a) and B(E2) rates (b) in $^{106}$Cd. The lines represents the calculated values using the classical particle plus rotor model. The solid and dashed lines in (a) are the calculated values for $V_{\pi \pi}$= 150 and 300 keV, respectively.}
\end{figure}

\begin{figure}
\includegraphics*[scale=0.7]{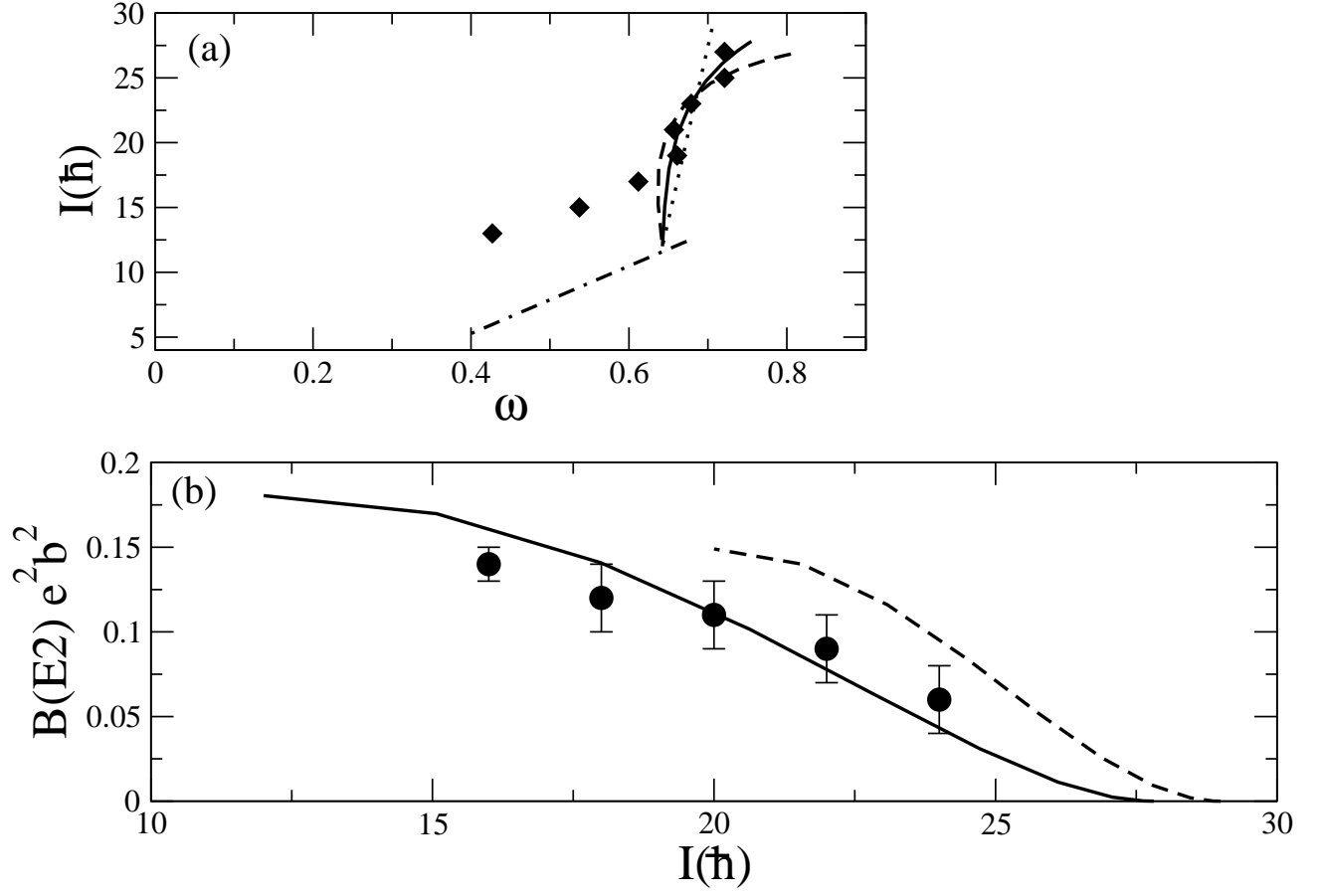}
\caption{\label{fig:fig6}~ The observed I($\omega$) plot (a) and B(E2) rates (b) in $^{110}$Cd. The dot-dashed line in (a) represents a rotor with moment of inertia of 19.2 MeV$^{-1} \hbar ^2$. The dotted, solid and dashed lines in (a) represents the calculated routhians for $V_{\pi \pi}$= 0, 150 and 300 keV, respectively. The solid and the dashed lines in (b) represents the calculated B(E2) values for AMR+rotation and pure AMR, respectively. }
\end{figure}

\newpage
\begin{table}
\caption{\label{tab:table1}~Measured level lifetimes and the corresponding B(E2) transition rates in $^{106}$Cd \cite{simons}, $^{108}$Cd \cite{pd1, simons2} and $^{110}$Cd. The error bars on the measured life-times include the fitting errors and errors in side-feeding intensities.}
\begin{ruledtabular}
\begin{tabular}{c|ccc|ccc|ccc}
&&$^{106}$Cd &&&$^{108}$Cd&&&$^{110}$Cd&\\
\hline
$I_i\rightarrow I_f$& $E_{\gamma}$ &$\tau$ &  $B(E2)$ & $E_{\gamma}$ &$\tau$ &  $B(E2)$& $E_{\gamma}$ &$\tau$ &  $B(E2)$ \\

 & (keV)&(ps)&$({eb})^2$& (keV)&(ps)&$({eb})^2$& (keV)&(ps)&$({eb})^2$\\

\hline

26$\rightarrow$24&1675.5&0.18$^a$&0.07$^a$ &&&&1443&0.35$^a$&-\\
24$\rightarrow$22&1487.6&0.19(02)&0.05(01)&&&&1358&0.27(07)&0.06(02)\\
22$\rightarrow$20&1310.6&0.26(02)&0.07(01)&1469&0.36 $^a$ &0.03$^a$ &1314&0.21(05)&0.09(02)\\
20$\rightarrow$18&1150.6&0.29(04)&0.14(02)&1260&0.28(03)&0.09(01)&1323&0.17(03)&0.11(02)\\
18$\rightarrow$16&980.8 &0.60(05)&0.15(01)&1105&0.33(03)&0.14(01)&1224&0.23(03)&0.12(02)\\
16$\rightarrow$14&&&&956&0.69(05)&0.15(01)&1075&0.36(03)&0.14(01)\\

\footnotetext[1]{Effective level lifetime.}
\end{tabular}
\end{ruledtabular}
\end{table}

\end{document}